\documentclass[aps,pra,reprint,showpacs]{revtex4-1}
\usepackage{graphicx}
\usepackage{bm}
\usepackage{bbm}
\usepackage{amsmath}
\DeclareMathOperator{\erf}{erf}
\DeclareMathOperator{\erfc}{erfc}
\DeclareMathOperator{\sign}{sign}
\renewcommand{\Im}{\operatorname{Im}}

\begin{document}

\title{Breakdown of atomic hyperfine coupling in a deep optical-dipole trap}

\author{Andreas Neuzner}
\author{Matthias K\"orber}
\author{Stephan D\"urr}
\author{Gerhard Rempe}
\author{Stephan Ritter}
\email{stephan.ritter@mpq.mpg.de}
\affiliation{Max-Planck-Institut f\"{u}r Quantenoptik, Hans-Kopfermann-Strasse 1, 85748 Garching, Germany}

\begin{abstract}
We experimentally study the breakdown of hyperfine coupling for an atom in a deep optical-dipole trap. One-color laser spectroscopy is performed at the resonance lines of a single $^{87}$Rb atom for a trap wavelength of 1064 nm. Evidence of hyperfine breakdown comes from three observations, namely, a nonlinear dependence of the transition frequencies on the trap intensity, a splitting of lines which are degenerate for small intensities, and the ability to drive transitions which would be forbidden by selection rules in the absence of hyperfine breakdown. From the data, we infer the hyperfine interval of the $5P_{1/2}$ state and the scalar and tensor polarizabilities for the $5P_{3/2}$ state.
\end{abstract}


\maketitle

Optical dipole traps (ODTs), including optical lattices, are established tools for trapping cold atoms. They rely on a position-dependent light shift of the atomic ground state. Excited states usually change differently so that transition lines are shifted. The line shifts can be small, like in shallow traps, or even vanish, like in magic-wavelength traps \cite{Katori1999, Hood:1999}. However, magic wavelengths cannot always be employed because either the spontaneous emission rate would be too high, or high-power lasers are not available, or more than two atomic levels are used, making it impossible to find a wavelength that is simultaneously magic for all transitions. Nevertheless, the increasing demand for improved control of atoms makes it desirable to work in deep ODTs. The advantages of deep ODTs include precise localization of the trapped atoms, the possibility to perform resolved-sideband Raman cooling, and reduced atom loss in the presence of heating processes.

Driving resonant transitions in deep ODTs requires precise knowledge of the atomic light shifts. For optical transitions, the excited-state light shifts are nontrivial and, interestingly, investigations on this subject are only at a beginning, although deep ODTs have long been employed in the fields of atomic clocks, quantum information processing (QIP), and quantum many-body physics. Examples include single-atom optical tweezers \cite{Tey2008d, Wilk2010, Isenhower2010, Hofmann2012b, Shih2013}, atoms inside high-finesse cavities \cite{Reiserer-arXiv:1412.2889}, atoms trapped close to nanophotonic waveguides \cite{Vetsch2010} or resonators \cite{Thompson2013}, atoms in hollow optical fibers \cite{Bajcsy2011}, and quantum gas microscopes \cite{Weitenberg2011, Parsons2015}. Line shifts might also become relevant in future experiments with ions in ODTs \cite{Enderlein2012, Linnet2012} and molecules in ODTs \cite{Ni2008, Danzl2010}.

Here we show that the differential light shifts depend nonlinearly on intensity for high intensities, in contrast to the well-known linear dependence for low intensities. The typical ODT depth, for which the nonlinear effect becomes comparable to the natural atomic linewidth is remarkably small, namely, $k_B\times 0.4$ mK for the parameters of our experiment, where $k_B$ is the Boltzmann constant. Moreover, we observe a splitting of resonances which would be degenerate in the linear regime and we observe a resonance which would be dipole forbidden in the linear regime. The data are obtained by varying the frequency of a probe laser near the $D_1$ or $D_2$ line, which illuminates a single $^{87}$Rb atom. When hitting a resonance, optical pumping transfers populations between hyperfine ground states. After a probe-light pulse, atomic populations are measured using cavity-enhanced state detection \cite{Bochmann2010}. From measured resonance positions at various ODT intensities, we infer the hyperfine splitting of the $5P_{1/2}$ state and the scalar and tensor polarizabilities at 1064 nm of the $5P_{3/2}$ state.

The physical reason for the observed nonlinear behavior is the breakdown of hyperfine coupling between nuclear spin and angular momentum of the valence electron \cite{Zon1973,Shore1981}. This is similar to the Paschen-Back effect, in which one observes breakdown of hyperfine or even fine-structure coupling when applying a strong static magnetic field. Hyperfine breakdown has also been observed in a static electric field \cite{Windholz1989}. Experimental evidence for hyperfine breakdown in a high-intensity light-field has been observed in a level-crossing spectroscopy experiment \cite{Gawlik:1991}, which used only one near-resonant light field that simultaneously caused and monitored hyperfine breakdown. To our knowledge, hyperfine breakdown in far-detuned ODTs has not been discussed in the literature.

We model the hyperfine breakdown largely analogous to the corresponding theory for an electrostatic field \cite{Windholz1989}. We approximate the Hamiltonian as $H= H_{\text{HF}}+H_\text{ODT}$. $H_{\text{HF}}$ describes the hyperfine interaction in the absence of external fields. It is diagonal in the basis of states $|J,I,F,m_F\rangle$, abbreviated as $|F,m_F\rangle$. Here, $J$, $I$, and $F$ are the angular-momentum quantum numbers of the valence electron, of the nuclear spin, and of the total atom. Their projections onto the $z$ axis are $m_J$, $m_I$, and $m_F$. The diagonal matrix elements $E_F$ of $H_{\text{HF}}$ in the $|F,m_F\rangle$ basis are independent of $m_F$. Throughout this work, we use the values measured in Ref.\ \cite{Ye1996} for the hyperfine intervals of the $5P_{3/2}$ state in $^{87}$Rb.

$H_\text{ODT}$ describes the effect of the ODT light onto the atom. The electric field of the ODT light at the atomic position is $\bm E(t)= \bm E_0 \cos(\omega t)$ with amplitude $\bm E_0$ and angular frequency $\omega$. We assume that the light is linearly polarized, and choose the $z$ axis along $\bm E_0$ ($\pi$ polarization). Considering the limit where all excited-state populations are small, the effect of the light is reduced to generating light shifts. In addition, considering the limit where all detunings are much larger than all hyperfine splittings, we approximate $H_\text{ODT}$ as being diagonal in the basis of states $|J,I,m_J,m_I\rangle$ (abbreviated as $|m_J,m_I\rangle$) with diagonal matrix elements
\begin{align}
E_{m_J}
= -\frac14 \left(\alpha_{\gamma}^S+\alpha_{\gamma}^T\frac{3m_J^2-J(J+1)}{J(2J-1)}\right) \bm E_0^2
\label{eqn:lightshift}
\end{align}
for $J\geq1$ and $E_{m_J}= -\frac14 \alpha_{\gamma}^S \bm E_0^2$ otherwise. This can be regarded as analogous to Ref.\ \cite{Windholz1989} combined with time averaging $\langle \bm E^2(t)\rangle_t =\bm E_0^2/2$. Note that these matrix elements are independent of $m_I$. Here, $\alpha_{\gamma}^S$ and $\alpha_{\gamma}^T$ are the scalar and tensor polarizabilities and $\gamma$ abbreviates the set of quantum numbers $n$, $L$, and $J$, where $n$ is the principal quantum number and $L$ the orbital angular momentum quantum number of the valence electron. An additional term containing the vector polarizability \cite{Arora2012} vanishes because we consider linear polarization. As the light is far detuned, the imaginary parts of the polarizabilities are negligible.

\begin{figure}[!tb]
\includegraphics[width=\columnwidth]{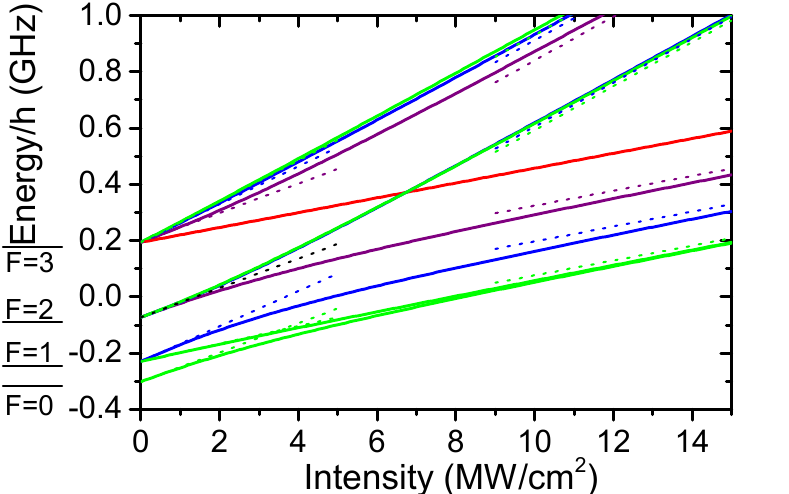}
\caption{(Color online) Energy levels of the $5P_{3/2}$ excited-state manifold in $^{87}$Rb in the presence of light shifts created by $\pi$-polarized 1064 nm light. Green, blue, purple, and red solid lines correspond to $|m_F|= 0$, 1, 2, and 3. At low intensity, $F$ is a good quantum number, whereas at high intensity $m_J$ becomes a good quantum number, similar to the hyperfine Paschen-Back effect. Dotted lines show linear approximations for low and high intensity. In the high-intensity region, the set of steeper lines correspond to $|m_J|= \frac12$. Note that an intensity of $1$ MW/cm$^2$ corresponds to a ground-state ODT depth of $k_B\times 1.5$ mK. The $|m_F|=0$ and 1 lines of the $F=2$ and $F=3$ manifolds are hardly resolved.
\label{fig:theory}
}
\end{figure}

In the low-intensity limit, $H_\text{HF}$ dominates and $H_\text{ODT}$ can be treated in first-order perturbation theory. Here, $H$ remains diagonal in the $|F,m_F\rangle$ basis. This yields low-intensity light shifts of the form $-\frac14 \alpha_{\gamma,F,m_F} \bm E_0^2 +\mathcal O(\bm E_0^4)$ with the low-intensity polarizability
\begin{align}
\alpha_{\gamma,F,m_F}
= \alpha_{\gamma}^S+\alpha_{\gamma,F}^T\frac{3m_F^2-F(F+1)}{F(2F-1)}
\label{eqn:approxpol}
\end{align}
for $F\geq1$ and $\alpha_{\gamma,F,m_F}= \alpha_{\gamma}^S$ otherwise. Here, $\alpha_{\gamma,F}^T= \alpha_{\gamma}^T \frac{3X(X-1)-4F(F+1)J(J+1)}{(2F+3)(2F+2)J(2J-1)}$ with $X= F(F+1)+ \linebreak[3] J(J+1)- I(I+1)$ \cite{Armstrong:1971}. For later reference, we note that this implies $\alpha_{5P3/2,F=2}^T= 0$ for $^{87}$Rb ($I=\frac32$). The emergence of $\alpha_{\gamma,F}^T$ from $\alpha_{\gamma}^T$ is somewhat reminiscent of the emergence of the hyperfine Land\'{e} factor $g_F$ from the fine structure Land\'{e} factor $g_J$.

\begin{figure}[!tb]
\includegraphics[width=\columnwidth]{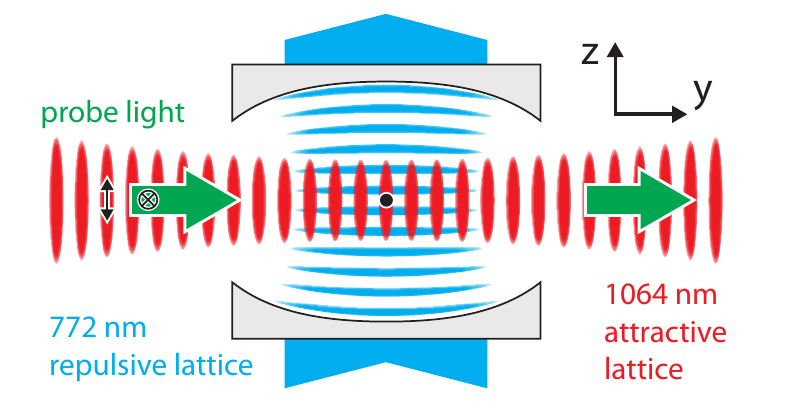}
\caption{(Color online) Scheme of the experimental setup. A single $^{87}$Rb atom is held in a two-dimensional optical lattice inside a Fabry-Perot resonator (gray). The relevant light shifts are created by the 1064 nm light. Probe light is applied for in-trap spectroscopy. The cavity and an additional laser (not shown) are used for hyperfine state detection.
\label{fig:setup}
}
\end{figure}

For intermediate intensities, the hyperfine coupling breaks down and $H$ is diagonal in neither the $|F,m_F\rangle$ nor the $|m_J,m_I\rangle$ basis. Here, $H$ can typically be diagonalized only numerically and the light shifts depend nonlinearly on $\bm E_0^2$. For high intensities (but not so high that fine-structure breakdown would occur), $H_\text{ODT}$ dominates and $H_\text{HF}$ can be treated in first-order perturbation theory. Here, $H$ is diagonal in the $|m_J,m_I\rangle$ basis and the energy eigenvalues are linear in $\bm E_0^2$ up to an offset.

\begin{figure*}[!t]
\includegraphics[width=\textwidth]{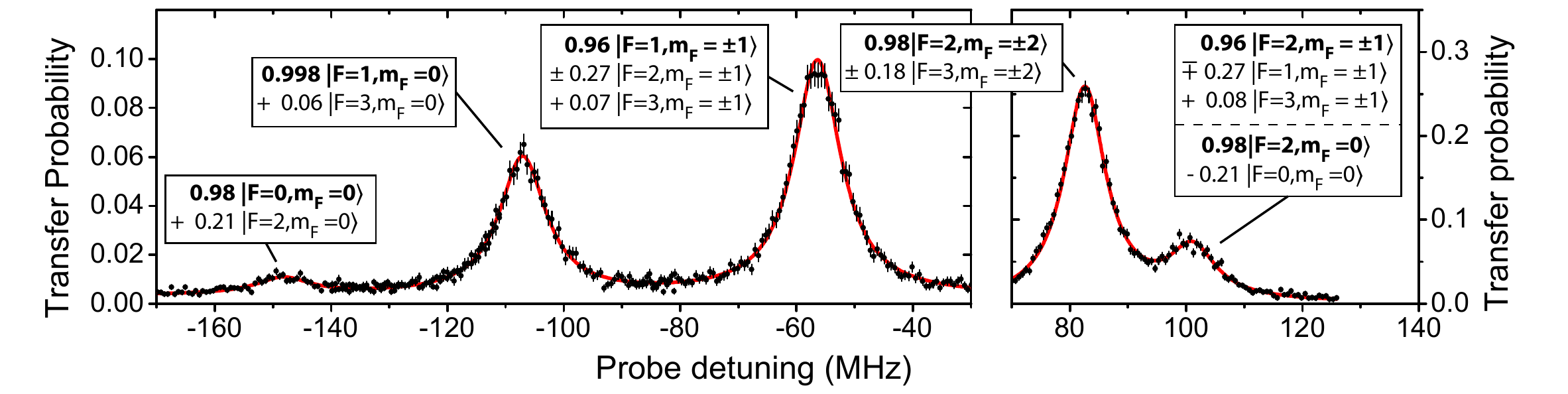}
\caption{(Color online) Measured excitation spectrum of a single $^{87}$Rb atom for transitions from the $5S_{1/2},F=1$ ground state to $5P_{3/2}$ excited states. A probe light pulse transfers population into the $F=2$ ground state by optical pumping, whenever resonant with an atomic transition. The final populations are measured. Five atomic resonance lines are clearly resolved. The line shows a fit of the sum of five Lorentzians with independent amplitudes, widths, and center frequencies. The boxes label energy eigenstates that we assign to the $5P_{3/2}$ excited states causing these resonances, where terms with amplitudes below 0.01 are omitted. Without hyperfine breakdown, this frequency range would show only three lines, because the two rightmost lines would be degenerate and the leftmost line would be a forbidden transition.
\label{fig:spectrum}
}
\end{figure*}

Figure \ref{fig:theory} shows the energy eigenvalues of $H$ for the $5P_{3/2}$ manifold of $^{87}$Rb as a function of the intensity $I_0= \frac12 c\varepsilon_0 \bm E_0^2$ of a plane traveling light wave, where $c$ is the vacuum speed of light and $\varepsilon_0$ the vacuum permittivity. This calculation is based on the theoretical predictions $\alpha_{5P3/2}^S= -1114(16)$ a.u.\ and $\alpha_{5P3/2}^T= 551(5)$ a.u.\ \cite{Safronova:pers} for 1064 nm [1 atomic unit $=1.648\,78\times10^{-41}$ J(V/m)$^{-2}$]. The figure shows the linear regimes for low and high intensities as well as the nonlinear behavior in the intermediate regime. Note that $\alpha_{5P3/2,F=2}^T= 0$ causes identical low-intensity behavior for the complete $F=2$ manifold, whereas different $|m_F|$ values within the $F=1$ or $F=3$ manifolds feature different initial slopes.

Figure \ref{fig:setup} shows a scheme of the experimental setup. A 772 nm light is coupled into the TEM$_{00}$ mode of a high-finesse Fabry-Perot resonator. This light creates a repulsive, one-dimensional (1D) optical lattice for ground-state $^{87}$Rb atoms with a potential height of $k_B\times 0.73$ mK. Additionally, a 1064 nm light beam crosses the resonator center. This beam has a waist ($1/e^2$ radius of intensity) of $w= 16$ $\mu$m and is $\pi$ polarized. It is retroreflected and creates an attractive, 1D lattice. A single $^{87}$Rb atom is loaded into the resulting two-dimensional (2D) lattice. The atom is located near an antinode of the 1064 nm light and near a node of the 772 nm light. Hence, at the position of the atom, light shifts from the 772 nm light are small (see the Appendix). The relevant light shifts for our measurements come from the 1064 nm light. Hopping of the atom between lattice sites is negligible.

An additional traveling-wave probe light beam with a wavelength near 780 nm or 795 nm crosses the cavity center perpendicularly to the cavity axis and subtending an angle of $45^\circ$ (out of the image plane in Fig.\ \ref{fig:setup}) with the 1064 nm light. This light is linearly polarized, perpendicularly to the $z$ axis, thus having equal fractions of $\sigma^+$ and $\sigma^-$ polarization. The probe light is frequency stabilized to a commercial frequency comb. All cavity resonances are far detuned from the probe light, so that the cavity does not affect the spectroscopy signal.

To perform spectroscopy, we first prepare the atom in the $F=1$ manifold of the $5S_{1/2}$ ground state with optical pumping using additional laser beams, not shown in Fig.\ \ref{fig:setup}. At this point, the atoms are approximately equally distributed among the $m_F$ substates of this $F=1$ manifold. Second, we expose the atom to a 2 $\mu$s long pulse of probe light at an intensity of roughly 0.6 mW/cm$^2$ and at a fixed frequency. Third, we measure whether the atomic population is still in the $F=1$ manifold or whether the probe light caused a spontaneous Raman transition into the $F=2$ manifold. This measurement uses cavity-enhanced state detection \cite{Bochmann2010}.

After loading the atom and after each such spectroscopy sequence, we perform polarization-gradient cooling using additional light beams to cool the atom to $\sim30$ $\mu$K. During this cooling process, we take $yz$-plane images of the cooling light scattered from the atom with a digital camera. With these images, we monitor whether the atom is still in place. A complete cycle for spectroscopy and cooling takes 0.5 ms. Such cycles are repeated for different frequencies of the probe light, until the atom is eventually lost. At that point, a new atom is loaded.

Figure \ref{fig:spectrum} shows a measured spectrum. The 1064 nm standing-wave ODT has a traveling-wave power of $P= 2.3$ W, corresponding to an ODT depth of $k_B\times 3.4$ mK. The data clearly show five resonance lines. The excited states belong to the $F=0$, 1, and 2 manifolds and are easily found in Fig.\ \ref{fig:theory}, in which these data correspond to 2.2 MW/cm$^2$. The widths of the resonances are $\sim1.5$ times the natural linewidth of the excited state. We attribute most of this broadening to the thermal position distribution of the atom (see the Appendix).

For the right part of Fig.\ \ref{fig:spectrum}, the hyperfine-coupled model Eq.\ \eqref{eqn:approxpol} predicts three degenerate lines for the $F=2$ manifold because $\alpha_{5P3/2,F=2}^T= 0$. The appearance of two resolved lines in this region unambiguously demonstrates hyperfine breakdown.

The leftmost line in Fig.\ \ref{fig:spectrum} belongs to the $F=0$ manifold. The $F=0$ excited state cannot decay to the $F=2$ ground state because of dipole-selection rules. Hence, this line would be invisible to our spectroscopy method according to the hyperfine-coupled model. But hyperfine breakdown gives the $F=0$ excited state an admixture of the $F=2$ exited state which can decay to the $F=2$ ground state, making the line visible. Note that a similar argument applies to all $F=3$ states.

\begin{figure}[!tb]
\includegraphics[width=\columnwidth]{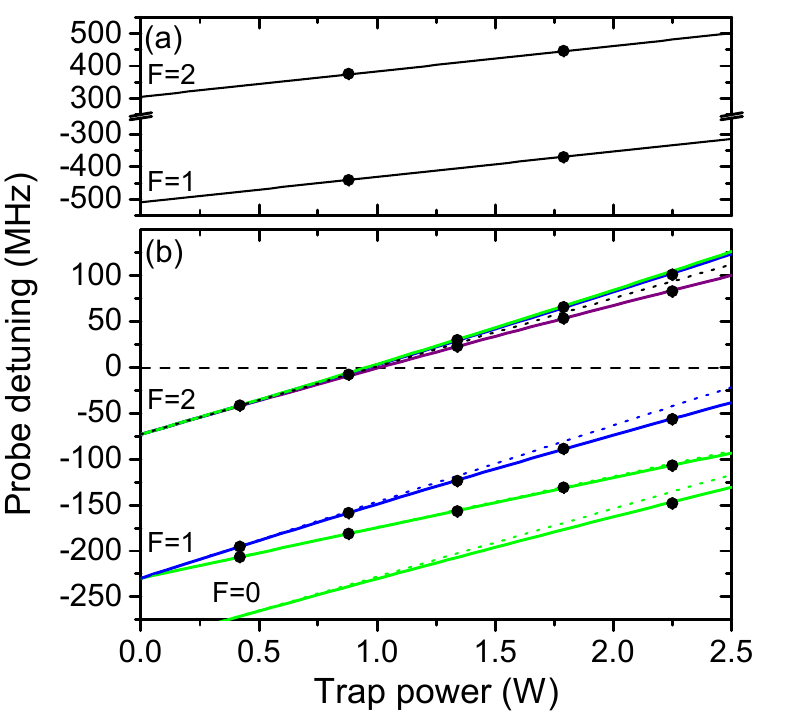}
\caption{(Color online) Quantitative analysis of differential light shifts. Solid lines show fits to the experimentally observed line centers (dots). Dotted lines show linear approximations for low power. The data cover the (a) $5S_{1/2}\leftrightarrow 5P_{1/2}$ and (b) $5S_{1/2}\leftrightarrow 5P_{3/2}$ transitions in $^{87}$Rb. Note that unlike Fig.\ \ref{fig:theory}, these data show differential light shifts.
\label{fig:ac-Stark-map}}
\end{figure}

Figure \ref{fig:ac-Stark-map} shows a quantitative analysis of the resonance positions. The values of the probe detunings at the line centers (dots) were extracted by fitting the sum of independent Lorentzians to spectra like in Fig.\ \ref{fig:spectrum} for different values of the traveling-wave power $P$ of the standing-wave ODT at 1064 nm. Statistical error bars are $\sim0.1$ MHz, much smaller than the symbol size. We label the detuning as zero, where the light is resonant with the transition from the $F=1$ ground state to the barycenter of the excited-state hyperfine manifold in the absence of light.

Figure \ref{fig:ac-Stark-map}(a) shows data for the $5S_{1/2}\leftrightarrow 5P_{1/2}$ transition. In the framework of Eq.\ \eqref{eqn:lightshift}, the tensor polarizabilities for both involved states vanish. Hence, we expect that the resonance positions depend linearly on $\bm E_0^2$ with a slope of $-\frac14\Delta\alpha_{1/2}$, independent of $F$. Here, we abbreviated $\Delta\alpha_J= \alpha_{5PJ}^S-\alpha_{5S1/2}^S$. In the experiment, there is some uncertainty in the power calibration, in the measured waist, and in the geometrical overlap of the incoming and retroreflected 1064 nm beams with the atom. We express this by replacing $\bm E_0^2\to \eta\bm E_0^2$ in the theory, introducing a dimensionless parameter $\eta$, which would ideally be unity. A fit to the data yields the best-fit values $\eta\Delta\alpha_{1/2}= -1675(3)$ a.u.\ and $(E_{F=2}-E_{F=1})/2\pi\hbar= 816.6(2)$ MHz. Only statistical uncertainties are quoted throughout this text. A thorough analysis of systematic uncertainties is beyond the scope of the present work. The value for $(E_{F=2}-E_{F=1})/2\pi\hbar$ agrees well with the measured value 816.66(3) MHz from Ref.\ \cite{Barwood1991}. This seems to create serious doubt in the measured value 812.29(3) MHz from Ref.\ \cite{Banerjee2004}, which deviates from our value by $\sim 20\sigma$.

Figure \ref{fig:ac-Stark-map}(b) shows data for the $5S_{1/2}\leftrightarrow 5P_{3/2}$ transition. A fit of the model based on Eq.\ \eqref{eqn:lightshift} to the data yields the best-fit values $\eta\Delta\alpha_{3/2}= -1590(3)$ a.u.\ and $\alpha_{5P3/2}^T/\Delta\alpha_{3/2}= -0.312(4)$. Combination with the best-fit value from Fig.\ \ref{fig:ac-Stark-map}(a) yields $\Delta\alpha_{1/2}/\Delta\alpha_{3/2}= 1.054(2)$.

We compare these results with the theory values $\alpha_{5S1/2}^S= 687.3(5)$ a.u.\ \cite{Arora2012}, $\alpha_{5P1/2}^S= -1226(18)$ a.u.\ \cite{Safronova:pers}, and the above theory values for $\alpha_{5P3/2}^{S/T}$, all for 1064 nm. They yield the theory predictions $\alpha_{5P3/2}^T/\Delta\alpha_{3/2}= -0.306(4)$ and $\Delta\alpha_{1/2}/\Delta\alpha_{3/2}= 1.062(14)$, in good agreement with our measured values.

As long as $\eta$ is unknown, our experimental data yield only ratios of polarizabilities. The simplest way to proceed is to assume $\eta=1$, which immediately yields experimental values for $\Delta\alpha_{1/2}$, $\Delta\alpha_{3/2}$, and $\alpha_{5P3/2}^T$. Another way to proceed is to use the above theory values for $\alpha_{5S1/2}^S$ and $\alpha_{5P1/2}^S$, which yields $\alpha_{5P3/2}^S= -1128(18)$ a.u., $\alpha_{5P3/2}^T= 568(9)$ a.u., and $\eta= 0.876(8)$. The latter reveals the size of the systematic uncertainty caused when one would set $\eta=1$.

Finally, we note that a measurement of the tensor polarizability $\alpha_{5P3/2}^T$ does not require the observation of hyperfine breakdown, because the light shifts in the low-intensity limit already reveal $\alpha_{5P3/2,F}^T$, from which one can easily calculate $\alpha_{5P3/2}^T$.

To conclude, we experimentally observed hyperfine breakdown induced by off-resonant light. This splits resonances, lifts selection rules, and creates shifts which are nonlinear in light intensity. The line shifts are relevant for precision experiments and high-fidelity QIP experiments, which drive resonant optical transitions in ODTs. The lifting of selection rules is relevant for all schemes that rely on the forbidden character of a transition, such as optical pumping schemes that populate an $m_F=0$ state. The observed effects should be even more dramatic for atoms with smaller hyperfine splittings, like lithium. For example, if a recent $^6$Li quantum gas microscope experiment \cite{Parsons2015} moved from the $D_1$ to the $D_2$ line, it would immediately operate deeply in the Paschen-Back regime.

A related experiment was simultaneously performed at the University of Michigan \cite{Chen-arxiv:1507.05675}.

\begin{acknowledgments}
We thank C.\ Hahn for contributions during an early stage of the experiment and M.\ Safronova for providing theory values for polarizabilities. This work was supported by the European Union (Seventh Framework Programme, Collaborative Project SIQS), by the Bundesministerium f\"{u}r Bildung und Forschung via IKT 2020 (Q.com-Q), and by Deutsche Forschungsgemeinschaft via NIM.
\end{acknowledgments}

\appendix*
\section{}

The spatial distribution of the atoms has a nonzero thermal width. As atoms at different positions experience different light shifts, the resonance lines in our experiment are broadened. As the broadening mechanism is not symmetric, this also causes a small shift of the line center. In the following, we describe a simple model for this broadening mechanism.

We can safely approximate the ground-state trapping potential $V_g(\bm x)$ as harmonic, because the atomic temperature is much smaller than the depth of the trapping potential. This yields $V_g(\bm x)= V_g(0) + U_g(\bm x)$ with $U_g(\bm x)= \sum_{i=1}^3 m\omega_i^2x_i^2/2$, where $m$ is the atomic mass and the $\omega_i$ are the angular frequencies of the ground-state trap. Similarly, we approximate the excited-state potential as $V_e(\bm x)= V_e(0) + U_e(\bm x)$ with $U_e(\bm x)= \sum_{i=1}^3 b_ix_i^2$, with certain coefficients $b_i$ which are typically negative in our experiment.

For a potential depth of $k_b\times 3.4$ mK of 1064 nm light and a potential height of $k_B\times 0.73$ mK of 772 nm light, we obtain trapping frequencies of $(\omega_x,\omega_y,\omega_z)/2\pi= (11,760,480)$ kHz. For $k_BT\gg \hbar\omega_i$ for all $i$ one could safely use a semiclassical approximation to calculate the thermal atomic position distribution. The typical atomic temperature in our experiment $T\sim30$ $\mu$K corresponds to $k_B T\sim 2\pi\hbar\times 600$ kHz so that we are near the border of the $k_BT\gg \hbar\omega_i$ regime. For simplicity, we nevertheless use a semiclassical approximation for the position distribution of the single atom
\begin{align}
\label{n(x)}
n(\bm x)
= n_0 e^{-\beta U_g(\bm x)}
,\end{align}
where $n_0$ is the peak density and $\beta= 1/k_BT$. The normalization condition $\int d^3x \, n(x)=1$ yields $n_0= (m\bar\omega^2/2\pi k_BT)^{3/2}$, where $\bar\omega=(\omega_x\omega_y\omega_z)^{1/3}$ is the harmonic mean of the trapping angular frequencies. Here, we assumed that different spatial directions are thermalized, which we did not verify experimentally. We measured the temperature only along the $z$ axis. For the density distribution \eqref{n(x)}, the standard deviations of the atomic position are $(\sigma_x,\sigma_y,\sigma_z)= (750,11,18)$ nm. Gravity points along the $x$ axis with a gravitational acceleration of $g=9.8$ m/s$^2$. This causes a gravitational sag of $\Delta x= g/\omega_x^2= 2$ nm, which is negligible compared to $\sigma_x$.

To see how this atomic position distribution affects the spectroscopy signal, we need to calculate how the number $N_2$ of atoms in the $F=2$ ground-state manifold changes in time because of optical pumping. Because the 2 $\mu$s probe pulse is much longer than the natural lifetime $1/\Gamma= 26$ ns of the excited state, we can use a rate-equation model to describe the temporal evolution of the atomic populations.

If the probe light near-resonantly drives the transition from a ground state $|g\rangle$ with $F=1$ and a given $m_F$ to an excited state $|e\rangle$, then the steady-state population in internal state $|e\rangle$ is
\begin{align}
\rho_{ee}(\bm x)
= \rho_{gg} \frac{I_p}{2I_s} \; \frac{1}{1+[2\Delta(\bm x)/\Gamma]^2}
,\end{align}
where $\rho_{gg}$ is the internal-state population in $|g\rangle$, $I_s$ the saturation intensity for the $|g\rangle\leftrightarrow |e\rangle$ transition, $I_p$ the probe-light intensity (for which we assumed $I_p\ll I_s$), and $\Delta(\bm x)= \omega_\text{probe}-\omega_\text{res}(\bm x)$ the detuning of the probe light from resonance, which depends on the position $\bm x$ because of the position-dependent differential light shifts
\begin{align}
\Delta(\bm x)
= \Delta_0 + \frac{U_g(\bm x)- U_e(\bm x)}\hbar
,\end{align}
where $\Delta_0= \Delta_\text{free}+V_g(0)-V_e(0)$ is the detuning at the trap center and $\Delta_\text{free}$ is the detuning for a free atom, in the absence of ODT light.

Let $\Gamma_2$ denote the partial rate coefficient for spontaneous radiative decay of state $|e\rangle$ into the $F=2$ ground-state manifold. Then averaging over the atomic density distribution yields
\begin{align}
\partial_t N_2
= \Gamma_2 \int d^3x \rho_{ee}(\bm x) n(\bm x)
.\end{align}

For simplicity, we restrict our considerations to a regime where the product $I_pt_p$ of the intensity $I_p$ and duration $t_p$ of the probe pulse is so low that we can approximate $\rho_{gg}$ as time independent. Hence, $\partial_t N_2$ is also time independent and  the quantity that we measure in our spectra is $N_2(t_p)= t_p \partial_t N_2$. As we are not particularly interested in the absolute heights of the spectral lines, it suffices to consider the dimensionless spectrum
\begin{align}
S(\Delta_0)
=  \frac{2I_s}{I_p\Gamma_2\rho_{gg}}\partial_t N_2
= n_0 \int d^3x \frac{e^{-\beta U_g(\bm x)}}{1+[2\Delta(\bm x)/\Gamma]^2}
,\end{align}
which is normalized to $\int S d\Delta_0 = \pi\Gamma/2$.

Using the harmonic approximation, we obtain
\begin{align}
S
= \frac1{\pi^{3/2}} \int d^3u \frac{e^{-u^2}}{1+4[\delta+(U_g-U_e)/\hbar\Gamma]^2}
,\end{align}
where we introduced a dimensionless detuning and the dimensionless coordinates
\begin{align}
\delta
= \frac{\Delta_0}{\Gamma}
,&&
u_i
= \sqrt{\frac{\beta m}2} \omega_i x_i
.\end{align}

In general, the dependence of the light shifts $V_{g/e}$ on trap intensity is nonlinear. However, for low enough atomic temperatures, the thermal atomic distribution samples only a small intensity range which is near the intensity at the trap center. Hence, we can safely approximate the light shifts $V_{g/e}(\bm x)$ as first-order Taylor polynomials in the intensity around the intensity at the trap center, yielding
\begin{align}
U_{g/e}(\bm x)
= \sum_{i=1}^2 -\frac14 \widetilde \alpha_{g/e,i} [\bm E_{0,i}^2(\bm x) - \bm E_{0,i}^2(0)],
\end{align}
with $\bm E_{0,i}(\bm x)$ associated with the $\lambda_i$ where $(\lambda_1,\lambda_2)= (1064,772)$ nm. Here we used that the trap intensity is proportional to $\bm E_0^2(\bm x)$. Our notation expresses the first-order Taylor coefficient in terms of $\widetilde \alpha_{g/e,i}$ which has the dimension of a polarizability and turns into $\widetilde \alpha_{g/e,i}= \alpha_{g/e,i}$ for the special case, where the light shift actually is linear in intensity.

For our experiment, the 1064 nm light provides the harmonic confinement along $x$ and $y$, whereas the 772 nm light provides the dominant harmonic confinement along $z$. In principle, the 1064 nm light provides additional harmonic confinement along $z$ but that is small and we neglect it. Hence
\begin{align}
\beta(U_g-U_e)
= \widetilde \alpha_{r,1} (u_x^2+u_y^2)+ \widetilde \alpha_{r,2} u_z^2
,\end{align}
where we abbreviated
\begin{align}
\widetilde \alpha_{r,i}
= 1- \frac{\widetilde \alpha_{e,i}}{\widetilde \alpha_{g,i}}
.\end{align}
In cylindrical coordinates, this yields the final result
\begin{align}
\label{S}
S
=
\frac{2}{\sqrt\pi} \int_0^\infty du_\varrho \int_{-\infty}^\infty du_z
\frac{u_\varrho e^{-u_\varrho^2-u_z^2}}{1+4(\delta+\tau_1 u_\varrho^2+ \tau_2 u_z^2)^2}
,\end{align}
where we introduced two dimensionless temperatures
\begin{align}
\tau_i
= \frac{k_BT}{\hbar\Gamma} \widetilde \alpha_{r,i}
.\end{align}
The integral \eqref{S} can be evaluated numerically.

\begin{table}[!b]
\caption{Polarizabilities in a.u.\ for $^{87}$Rb.
\label{tab:alpha}}
\begin{tabular*}{\columnwidth}{c@{\extracolsep\fill}cccc}
\hline\hline
$\lambda$ & $\alpha_{5S1/2}^S$ & $\alpha_{5P1/2}^S$ & $\alpha_{5P3/2}^S$ & $\alpha_{5P3/2}^T$ \\
\hline
1064 nm & $687.3(5)$ & $-1226(18)$ & $-1114(16)$ & $551(5)$ \\
772 nm & $-11995$ & $1450$ & $3460$ & $-4842$ \\
\hline\hline
\end{tabular*}
\end{table}

Table \ref{tab:alpha} shows theory values for the scalar and tensor polarizabilities relevant for our experiment. The values for 1064 nm are the theory values from our paper. We calculated the values for 772 nm ourselves in second-order perturbation theory, including transitions $5S_{1/2}\leftrightarrow n'P_{J'}$ and $5P_J\leftrightarrow n'S_{1/2}$, and $5P_J\leftrightarrow n'D_{J'}$ using resonance frequencies and oscillator strengths from Refs.\ \cite{NIST:atomic-spectra-database, Safronova:11, Arora:07}.

\begin{table}[!b]
\caption{Values of $\alpha_{r,J}^{S/T}= 1-\alpha_{5PJ}^{S/T}/\alpha_{5S1/2}^S$ for $^{87}$Rb based on Table \ref{tab:alpha}.
\label{tab:ratios}}
\begin{tabular*}{\columnwidth}{c@{\extracolsep\fill}ccc}
\hline\hline
$\lambda$ & $\alpha_{r,1/2}^S$ & $\alpha_{r,3/2}^S$ & $\alpha_{r,3/2}^T$ \\
\hline
1064 nm & $2.78$ & $2.62$ & $0.20$ \\
772 nm & $1.12$ & $1.29$ & $0.60$ \\
\hline\hline
\end{tabular*}
\end{table}

Table \ref{tab:ratios} lists the quantity $\alpha_{r,J}^{S/T}= 1-\alpha_{5PJ}^{S/T}/\alpha_{5S1/2}^S$, which is related to  $\widetilde \alpha_{r,i}$, calculated from Table \ref{tab:alpha}. Of course, one could calculate Eq.\ \eqref{S} numerically for each excited state. Here, however, we consider only one example, namely, for the typical values $\widetilde \alpha_{r,1}= 2.7$ for 1064 nm and $\widetilde \alpha_{r,2}= 1.2$ for 772 nm. Combination with $k_BT/\hbar\Gamma= 0.10$ yields
\begin{align}
\label{tau}
\tau_1
= 0.27
,&&
\tau_2
= 0.12
.\end{align}

\begin{figure}[!t]
\includegraphics[width=\columnwidth]{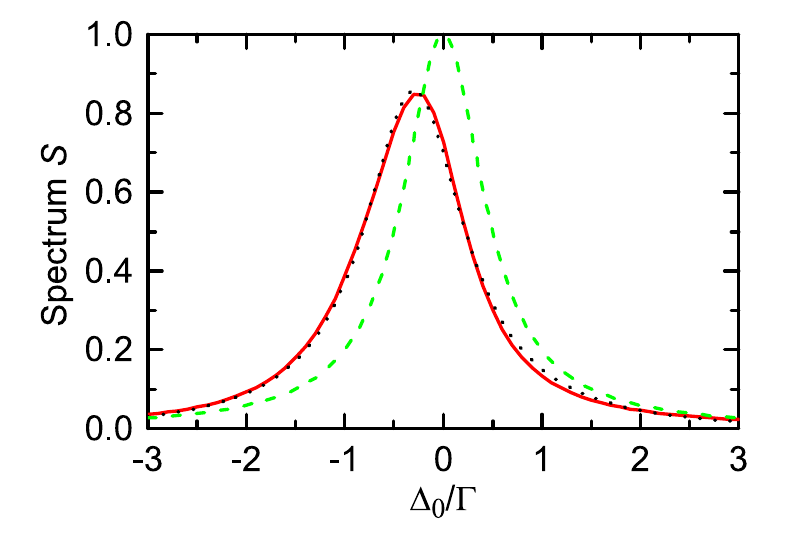}
\caption{(Color online) Expected spectral line shapes. The red solid line shows the expected line shape numerically calculated from Eq.\ \eqref{S} for the typical values for our experiment from Eq.\ \eqref{tau}. This line is broadened and shifted because of the nonzero width of the thermal atomic position distribution. The black dotted line is a Lorentzian fit to this. The green dashed line shows the Lorentzian reference for $\tau_1=\tau_2=0$, corresponding to a classical particle located exactly at the trap center.
\label{fig:spectral-line-shape}}
\end{figure}

In Fig.\ \ref{fig:spectral-line-shape}, numerical results for the integral \eqref{S} for these values of $\tau_1$ and $\tau_2$ are shown as a red solid line. The black dotted line shows a fit of a Lorentzian to this result. The best-fit values are a full width at half maximum (FWHM) of $1.25\Gamma$ and a center position of $-0.30\Gamma$. For reference, the green dashed line shows the trivial Lorentzian obtained for $\tau_1= \tau_2= 0$, which has a FWHM of $\Gamma$ and is centered at 0. The latter can be regarded as the zero-temperature limit of Eq.\ \eqref{S}, although in this limit, the semiclassical approximation is expect to be poor.

Comparison of the measured line widths of $\sim 1.5 \Gamma$ with the width $1.25\Gamma$ from the fit suggests that the line broadening observed in the experiment is to a good part caused by the thermal atomic position distribution.

We emphasize that the line shifts and line broadenings in this model are independent of the depth of the 1064 nm and height of the 772 nm potential, as long as the traps are strong enough that the harmonic approximation works well and as long as they are weak enough that the semiclassical approximation works well.

The line shift found in this model suggests that all lines measured in our experiment are systematically red shifted by $\sim 0.3\Gamma \sim 1.8$ MHz relative to the free-space resonance lines. Note that within the approximations used so far, this has no effect on the polarizabilities and hyperfine splittings that we extract from the data. Those values will only be affected if additional corrections are taken into account, such as the fact that the actual values of $\widetilde \alpha_{r,i}$ are somewhat different for different atomic states. Because of the lack of a tensor polarizability in the $5S_{1/2}\leftrightarrow 5P_{1/2}$ transition, however, this is not the case for values extracted from those data, including the $5P_{1/2}$ hyperfine splitting.

To experimentally test the sensitivity to a change in the 772 nm potential height, we measured the resonance line involving the $F=|m_F|=1$ excited state at two potential heights, namely $k_B\times0.55$ and 1.1 mK. The relative shift of the line center between these two experiments was measured to be 79(91) kHz, which is consistent with zero.

Finally, we note that if only one wavelength was used for the optical lattice, the integral \eqref{S} could be solved analytically. To this end, one could rewrite it in spherical coordinates as
\begin{align}
S
= \frac{4}{\sqrt\pi} \int_0^\infty du_r \frac{u_r^2 e^{-u_r^2}}{1+4(\delta+\tau_1 u_r^2)^2}
.\end{align}
For $\delta,\tau_1\in\mathbbm R$ with $\tau_1\neq0$, this integral has the analytic solution
\begin{align}
\label{s-analytic}
S
= \frac{\sqrt{\pi}}{2\tau_1} z w(z) + \text{c.c.}
,&&
z=
\sqrt{\frac{i-2\delta}{2\tau_1}} \sign (\tau_1)
,\end{align}
where \cite{abramowitz:72}
\begin{align}
w(z)
= e^{-z^2}\erfc(-iz)
\end{align}
is the Faddeeva function, $\erfc(z)= 1-\erf(z)$ the complementary error function, and $\erf(z)= \frac{2}{\sqrt\pi} \int_0^z e^{-t^2}dt$ the error function. Equation \eqref{s-analytic} can be derived using partial-fraction decomposition and the integral representation \cite{abramowitz:72}
\begin{align}
\label{abramowitz}
w(z)
= \frac{2iz}{\pi} \int_0^\infty \frac{e^{-t^2}}{z^2-t^2}
,&& \Im z>0
.\end{align}

Note that Eq. \eqref{abramowitz} can also be used to analytically solve the $u_z$ integral in \eqref{S}, yielding

\begin{align}
S
=\frac{2}{\sqrt{\pi}}\int_0^{\infty}du_{\rho}e^{-u_{\rho}^2}u_{\rho}I_z(u_{\rho})
\end{align}
with
\begin{align}
I_z=\frac{\pi}{4\tau_2 z}w(z)+c.c.
,&&
z=\sqrt{\frac{i-2\delta-2\tau_1u_{\rho}^2}{2\tau_2}}\sign(\tau_2)
\end{align}
~~

\end{document}